\documentclass[runningheads]{llncs}
\usepackage[T1]{fontenc}
\usepackage{graphicx}
\usepackage{algorithm}
\usepackage{algpseudocode}

\def\BibTeX{{\rm B\kern-.05em{\sc i\kern-.025em b}\kern-.08em
    T\kern-.1667em\lower.7ex\hbox{E}\kern-.125emX}}

\usepackage{CJKutf8}

\usepackage{xcolor}
\usepackage{pgfplots}
\usepackage{tikz}
\usepackage{comment}
\usepackage{pgfplotstable}

\usepackage{filecontents}

\usepackage{pifont} 
\usepackage{amssymb}
\usepackage{bbding} 
\usepackage{pifont}

\usepackage{makecell}
\usepackage{colortbl} 

\usepackage{fancyhdr} 
\usepackage{url}

\usepackage{multirow}
\usepackage{cite}
\usepackage{amsmath,amssymb,amsfonts}

\usepackage{textcomp}
\usepackage{xcolor}

\usepackage{subfigure}
\usepackage{pgfplots}
\usepackage{filecontents}
\usepackage{pgf-pie}

\usepackage{graphicx,subfigure}

\definecolor{bblue}{HTML}{4F81BD}
\definecolor{rred}{HTML}{C0504D}
\definecolor{ggreen}{HTML}{9BBB59}
\definecolor{ppurple}{HTML}{9F4C7C}

\usepackage{listings}

\usepackage{blindtext}
\usepackage{etoolbox,xstring,mfirstuc,textcase}
\usepackage{booktabs}
\usepackage{pgfplots}

\usepackage{tabularx,booktabs,makecell,multirow, caption}
\usepackage{enumitem} 

\newcolumntype{L}{>{\arraybackslash}X}

\begin{document}
\title{Are NFTs Ready to Keep Australian \\ Artists Engaged?
}

\author{Ruiqiang Li\inst{1}
\and
Brian Yecies\inst{1}
\and
Qin Wang\inst{2}
\and 
Shiping Chen\inst{2}
\and
Jun Shen\inst{1}
}
\authorrunning{R. Li et al.}

\institute{University of Wollongong, Australia \and
CSIRO Data61, Australia}

\maketitle 
\setcounter{footnote}{0}

\begin{abstract}
Non-fungible Tokens (NFTs) offer a promising mechanism to protect Australian and Indigenous artists'
copyright. They represent and transfer the value of artwork in digital form. Before adopting NFTs to protect Australian artwork, we in this paper investigate them empirically. We focus on examining the details of NFT structure. We start from the underlying structure of NFTs to show how they represent copyright for both artists and production owners, as well as how they aim to safeguard or secure the value of digital artworks. We then involve data collection from various types of sources with different storage methods, including on-chain, centralized, and decentralized systems. Based on both metadata and artwork content, we present our analysis and discussion on the following key issues: copyright, security and artist identification. The final results of the evaluation, unfortnately, show that the NFT is NOT ready to protect Australian and Indigenous artists' copyright. 
\keywords{Blockchain  \and Non-Fungible Token \and Copyright \and Security.}
\end{abstract}
%
%
%
\section{Introduction}

Australia has one of the largest visual art markets, valued at \$1.8 billion~\cite{AACHWA_PricingArtMarket}, but it is currently facing significant issues with copyright infringement. The Arts Law Centre of Australia has also reported a concerning rise in such cases in recent years \footnote{Copyright Enforcement Review Submission: \url{https://www.artslaw.com.au/wp-content/uploads/2023/03/Arts-Law-Centre-of-Australia-Submission-on-Copyright-Enforcement-Review-2022-23-7-March-2023.pdf}}. Moreover, copyright violations have also affected Indigenous artists \footnote{An Indigenous artist discovers his rights are infringed multiple times: \url{https://www.artslaw.com.au/case-studies/indigenous-artist-discovers-his-rights-infringed/}}. To better protect Australian artworks, we consider employing Non-Fungible Token (NFT) technology~\cite{wang2021non,li2024empowering}, which offers traceability and automated royalty distribution. However, before adopting this technology, it is important to investigate the current status of NFTs from their underlying structure to explore how they support copyright protection and the security of NFTs. Furthermore, we explore whether it is feasible to identify Australian artists through NFT metadata, which could assist the Australian Copyright Agency in collecting and managing artists’ works on-chain.

NFTs had emerged as a revolutionary, yet promising, technology in the digital art and collectibles space, offering a unique way to authenticate, own, and trade digital assets. Unlike traditional cryptocurrencies, which are fungible and interchangeable, NFTs are unique digital tokens that represent ownership of a specific item, such as artwork, music, or virtual real estate. At the core of NFTs lies the metadata, which serves as the backbone of these tokens by providing essential information about the asset, including its name, description, and link to the associated artwork or file. This metadata is crucial for defining the identity and value of an NFT, making it a key area of study for understanding the broader implications of this technology.

This paper focuses on the metadata of NFTs built on the case of emerging ERC-721 protocol, one of the most widely adopted standards for creating and managing NFTs on the Ethereum blockchain. The ERC-721 standard has become a cornerstone of the NFT ecosystem, enabling the creation of unique tokens with distinct properties and metadata. By examining the metadata of ERC-721-based NFTs, this study aims to provide a detailed analysis of how metadata is stored, structured, and utilized in practice. Specifically, it explores the different storage methods for metadata, such as on-chain and off-chain storage, and evaluates their implications for copyright, security, and cost. Additionally, the study investigates the structure of metadata, including the types of information it contains and how this information is organized to support the functionality and value of NFTs.

Another critical aspect of this paper is the examination of how artwork associated with NFTs is stored and accessed. While the metadata provides a link to the artwork, the actual storage of the artwork itself can vary significantly, ranging from decentralized storage solutions like IPFS to centralized cloud services. This variation raises important questions about the security, and accessibility of the artwork, which are essential considerations for both creators (e.g., artists) and collectors (e.g. art piece owners). Furthermore, this paper explores the potential for artist identification through metadata analysis. By examining the metadata, it might be possible to trace the origin of an NFT and verify its creator, which could play a crucial role in addressing copyright issues and ensuring the authenticity of digital assets.

Our analysis also delves into the challenges and opportunities associated with NFT metadata, particularly in the areas of copyright, and security. Copyright issues are a significant concern in the NFT space, as the metadata and associated artwork can be easily copied or misrepresented, leading to disputes over ownership and intellectual property rights. Security issues, such as vulnerabilities in metadata storage~\cite{stoger2023demystifying} or the potential for malicious manipulation~\cite{abidin2013printmaking}, further complicate the landscape. By exposing these issues, we aim to provide a comprehensive understanding of the technical, legal, and practical implications of NFT metadata, offering insights into its role in the rapidly evolving digital economy. 

\begin{figure}[htp]
    \vspace{-20pt}
    \centering
    \centerline{\includegraphics[width=0.9\columnwidth]{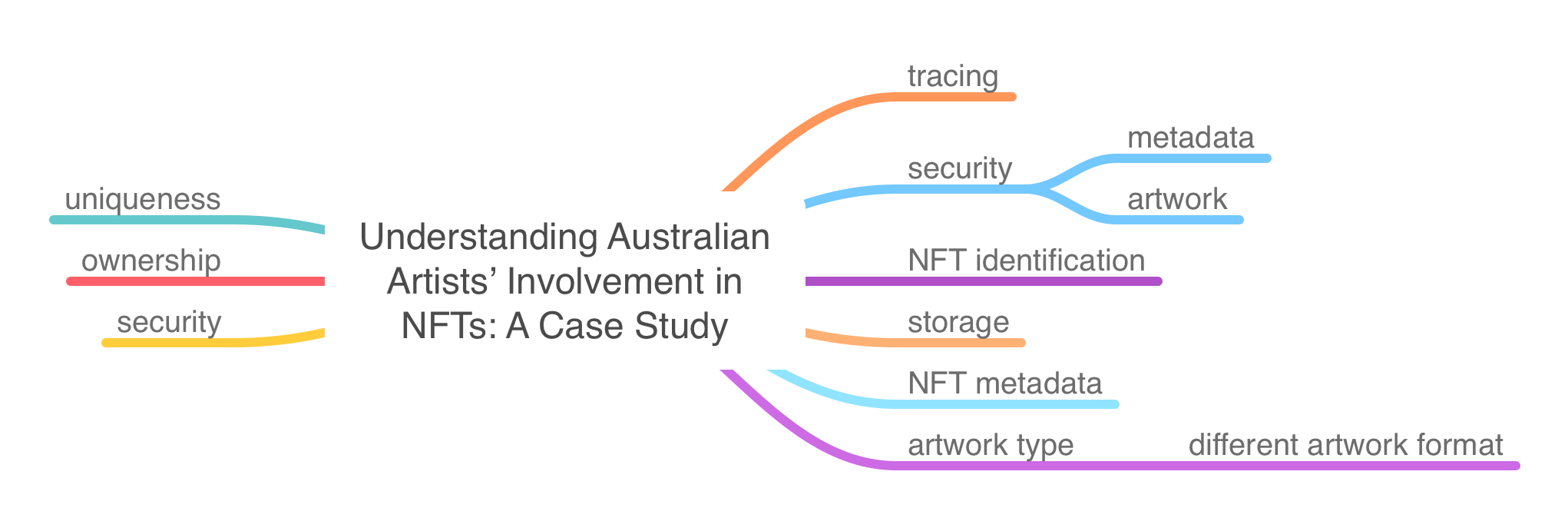}}
    \caption{Study Map}
    \label{fig:map1}
    \vspace{-10pt}
\end{figure}
In this paper we try to examine and answer the following research questions:
\begin{itemize}
    \item \textbf{RQ1. What kinds of copyright issues arise from NFT metadata and artwork?}
    \item \textbf{RQ2. What kinds of security issues arise from NFT metadata and artwork?}
    \item \textbf{RQ3. Is it easy to identify artists from the metadata level?}
\end{itemize}

We have the following contributions (Figure \ref{fig:map1} showing our roadmap):
    \vspace{-3pt}
\begin{itemize}
    \item We collect NFT metadata and artwork data from on- and off-chain storage. 
    \item We analyse copyright issues and security issues for NFTs based on metadata and artwork.
    \item We carry out the Australian artist's identification from NFT metadata.
\end{itemize}

\section{Quick Access to NFT}

\begin{figure}[htbp]
    \vspace{-15pt}
    \centering
    \centerline{\includegraphics[width=0.9\columnwidth]{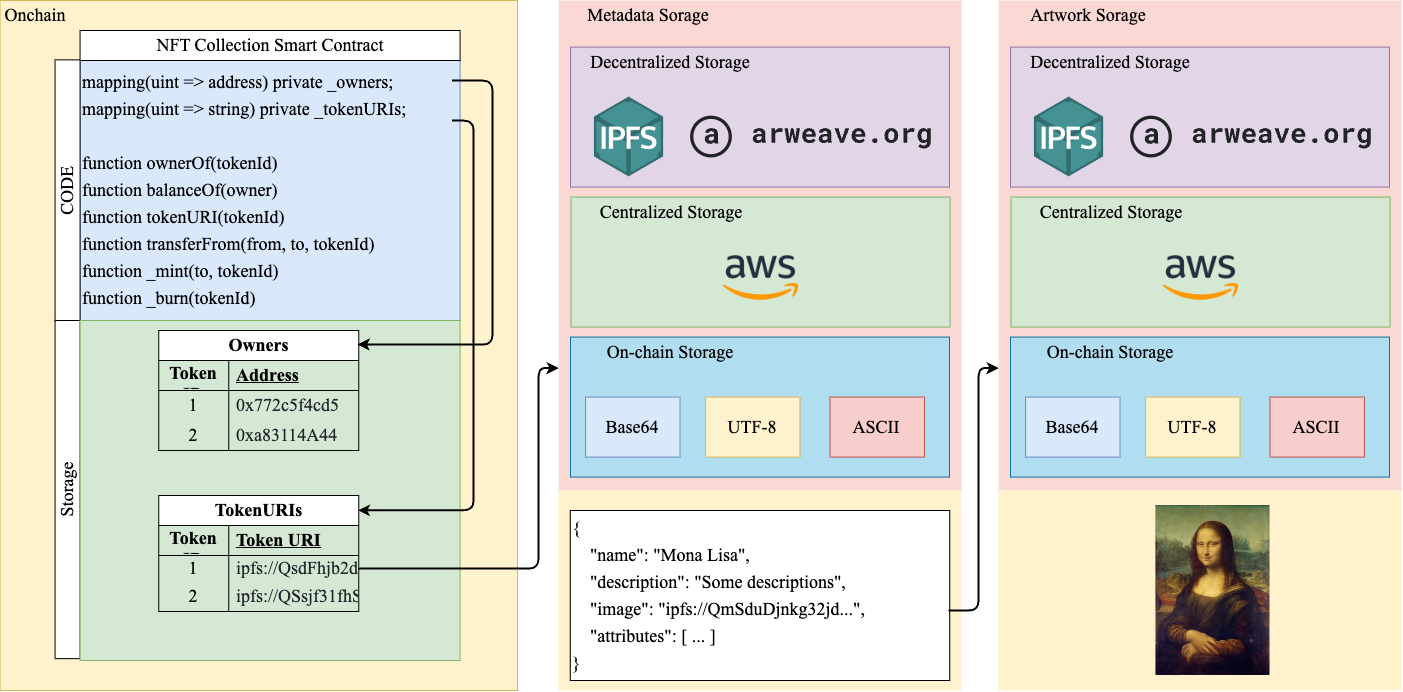}}
    \caption{NFT Research Focus}
    \label{fig:NFT Research Focus}
    \vspace{-15pt}
\end{figure}

NFTs are digital assets stored on a blockchain that represent ownership of unique items, such as artwork, music, or virtual goods. The structure of an NFT can be divided into three key parts: the on-chain contract, metadata storage, and artwork storage. Each component plays a critical role in defining the functionality, ownership, and presentation of the NFT. The on-chain component includes crucial information such as the NFT contract address, which serves as the unique identifier for the collection of tokens deployed on the blockchain. Each NFT within the collection is further identified by its unique tokenID. Ownership information is recorded on-chain, linking the NFT to its current owner's wallet address. The \emph{tokenURI} is also stored on-chain, which serves as a pointer to the metadata that describes the NFT. This URI is essential as it connects the on-chain record to oﬀ-chain metadata. Metadata storage is where the descriptive information about the NFT is stored. The metadata is typically in JSON format and includes fields such as name, description, image, and other attributes like traits or rarity information. Metadata can be stored in decentralized systems like IPFS (InterPlanetary File System) or Arweave, which offer enhanced security and permanence. Alternatively, centralized solutions like Amazon Web Services (AWS) are also commonly used for their ease of use and scalability, though they rely on third-party servers. Some metadata is even stored directly on-chain, using encoding methods such as BASE64, UTF-8, or ASCII. While this provides maximum immutability, it is typically reserved for high-value NFTs due to its high storage costs. Artwork storage refers to where the actual visual or audio representation of the NFT resides. The same storage types used for metadata are also applicable here including decentralized storage (IPFS, Arweave), centralized storage solutions, such as Amazon and other cloud providers and on-chain storage of artwork encoded using BASE64, UTF-8, or ASCII. By combining these components, NFTs ensure a ﬂexible and secure framework for representing ownership and uniqueness. Figure \ref{fig:NFT Research Focus} illustrates the entire clear NFT structure with various storage methods.
\vspace{-10pt}
\subsection{Copyright}
Recent research on NFTs has increasingly focused on copyright protection, ownership verification, and secure transactions. Wang et al. propose referable NFTs (rNFTs) using a DAG structure to enhance visibility and profit-sharing among creators~\cite{wang2023referable,yu2025maximizing}. Kimura et al. address risks from blockchain forks with a cross-referencing scheme to preserve NFT uniqueness and platform linkage~\cite{kimura2023cross}. Kotzer et al. explore image hash functions to detect duplicated NFTs, a key concern for visual copyright enforcement~\cite{kotzer2024detection}. Igarashi et al. present Photrace, a blockchain-based system for authenticating image provenance via secure digital camera certificates~\cite{igarashi2021photrace}. Liu et al. propose encryption and watermarking techniques to protect NFT artworks from unauthorized use, highlighting the need for both privacy and secure circulation\cite{liu2024privacy}. Lastly, Maesa et al. introduce Non-Fungible Mutable Tokens (NMTs) for dynamic digital assets, emphasizing access control to prevent unauthorized changes~\cite{maesa2024protecting}. Together, these studies offer diverse strategies for safeguarding NFT copyright in evolving digital environments.
\vspace{-10pt}

\subsection{Security}
The security of NFTs is a growing concern as studies highlight various technical and systemic vulnerabilities. Mochram et al. identify issues like copyright infringement, data theft, and plagiarism that blockchain alone cannot solve, calling for clearer ownership definitions and proactive protection by users~\cite{mochram2022systematic}. Stöger et al. expose how centralized components in Web3 can lead to NFT hijacking, challenging the decentralized promise of blockchain~\cite{stoger2023demystifying}. Wang et al. reveal the fragility of the NFT-to-asset link due to poor decentralized storage and data duplication~\cite{wang2023nfts}, while Das et al. point to marketplace design flaws that expose users to scams~\cite{das2022understanding}. Ma et al. detect widespread smart contract defects with their tool Emerium~\cite{ma2025uncovering}, and Li et al. highlight risks from metadata tampering and invalid caching, which can break the link between NFTs and their provenance~\cite{li2024empowering}. Together, these studies underscore the need for stronger technical safeguards, better infrastructure, and user awareness to secure the NFT ecosystem.

Current research primarily focuses on smart contracts, such as NFT smart contracts and NFT marketplace contracts, and artwork. However, less attention is paid to NFT metadata and the relationships between NFT smart contracts and metadata, as well as between metadata and the associated artwork. NFT metadata and artwork themselves involve copyright issues, security issues, and challenges in artist identification. This paper aims to expose these issues using real NFT data and provide some insight for future research.

\section{Methodology and Experiment}
\label{sec-method}

The NFT contract addresses and on-chain data, including encoded metadata or metadata URLs, are collected from the Ethereum Mainnet RPC, which is the endpoint for Ethereum Mainnet and the channel for data between outside and Ethereum inside. From the \emph{tokenURI} in each NFT contract, we locate the metadata storage and retrieve the metadata. Since the metadata is encoded using various methods, it needs to be decoded after collection. Within the metadata, while the image URL points to the location where the associated artwork is stored. Additionally, information about copyright and licensing can be found in the metadata. It is also used to identify Australian artists from this metadata.

The dataset has 768,037 contract addresses from Block 16,000,000 to Block 21,269,338. 43,810 ERC721 tokens are identified from the contract address. 

Our experiment hardware inclcues CPU: Intel(R) Xeon(R) CPU E5-2690 v4 @ 2.60GHz *2, Memory: 256 GB, NVME: 8 TB. The experiment software includes Erigon Version: v2.59.2.

\smallskip
\noindent\textbf{Collecting NFT contract addresses.}
The Ethereum Mainnet RPC endpoint is self-built by the Erigon \footnote{https://github.com/erigontech/erigon} locally, which can provide unlimited usage and high data-collecting speed. The NFT contract addresses cannot be acquired from on-chain data directly; they need to be identified with different token types. There are 4 token types: ERC20, ERC721, ERC777, ERC1155. The ABI interface is used to identify the token type. The interfaces for the different tokens are different; we further construct the instance for the token with its interface and smart contract address, and then try to call the method of the token. If the call is successful, the token type is found. Otherwise, the contract is not the target.

\smallskip
\noindent\textbf{Collecting NFT metadata.}
Most NFT metadata is stored in off-chain storage, such as decentralized storage and cloud storage. However, some NFT metadata is stored on-chain as a smart contract, which is encoded with different encoding methods, such as Base64, UTF-8, and ASCII. If the data from the \emph{tokenURI} interface is not the correct URL, it is encoded metadata. For off-chain storage, the metadata can be acquired by sending request to the token URI. However, for the decentralized storage, such as IPFS and Arweave, some prefixes of URIs need to be updated with third-party providers’ services because the normal IPFS and Arweave URIs with prefixes \texttt{ipfs://} and \texttt{ar://} cannot be requested correctly.  For on-chain storage, the metadata can be acquired with decoding with different encoding methods by the prefix in the metadata. For example, the prefix \texttt{data:application/json;base64} means this metadata is encoded with Base64. The prefix \texttt{data:application/json;utf-8} means this metadata is encoded with UTF-8, and the prefix \texttt{data:application/json;ascii} means the metadata is encoded with ASCII. Most NFTs utilize the metadata standard from OpenSea, which includes the image URL as one attribute within the metadata\footnote{OpenSea Metadata Standards: \url{https://docs.opensea.io/docs/metadata-standards}}. This metadata is formatted in JSON\footnote{JSON: \url{https://www.json.org}}, facilitating the extraction of the image URL from the image attribute.

\smallskip
\noindent\textbf{Collecting NFT artwork.}
Artworks are predominantly stored in off-chain storage solutions, encompassing both decentralized storage and cloud storage. Conversely, some artworks are stored on-chain using various encoding methods, such as Base64 and UTF-8. These on-chain artworks are prefixed differently, for example, \texttt{data:image/<image type>;base64} or \texttt{data:image/<image type>;utf8}. The methods employed to collect artwork mirror those used for metadata collection.

\smallskip
\noindent\textbf{Collecting NFT copyright and license information.}
The copyright and license information are all in the metadata. We need to search keywords `copyright' and `license' in the metadata to identify which NFT has copyright information first. The keywords `copyright' and 'license' are in four different locations. First: the copyright or license is an attribute in metadata. Second, the copyright or license is in the description of metadata. Third, the copyright or license is under attributes that follow the Opensea metadata standards. Fourth, the copyright is under the license attribute. Furthermore, we also need to extract copyright information from the NFT metadata which has keyword `copyright' or `license'. There are some standard licenses, such as CC0, CBE-CC0, CC BY-NC-ND 4.0, CC-BY-4.0, CC BY-NC-SA 4.0, CC BY-NC 4.0, CC BY NA 2.0, CC BY-NC 2.0, Art Tokyo Global\footnote{Art Tokyo Global
: \url{https://hub.xyz/arttokyoglobal}}. However, some copyright information does not include the standard license. After searching keywords `copyright' and `license' from NFT metadata, we find that the NFT metadata has one keyword but NFT metadata does not have any copyright or license information. This type of NFT metadata could not be identified as NFT including copyright information and they are removed from the final collection.

\vspace{-10pt}
\subsubsection{Australian artists identification.}
The metadata often contains the author’s name and description of the artwork. The intuitive approach to identify Australian artists is to search artists' names in the metadata which is a string format from converting JSON to string. The Australian artist's name list is provided by the Australian Copyright Agency \footnote{Australian Copyright Agency: \url{https://www.copyright.com.au}}. Moreover, the Australian artist's name list is processed by removing the middle name, because the author name in NFT metadata does not have the middle name, either. For accuracy consideration, the full name and that without middle name are both used as keywords to search in metadata. Another way to identify artists is to search 'Australia' as a keyword in metadata. It is the same way as searching an artist's name in metadata.

\section{Results}
\label{sec-result}

We collected a total of 13,462 correct metadata from the identified ERC721 tokens. From metadata, we extracted the image and found 9 formats of the image URL correspondigly. We also serached the artist's name, which yielded 199 Australian artists. We give the breakdown details for those infomation.

\begin{figure}[htp]
\vspace{-5pt}
    \centering
    \centerline{\includegraphics[width=0.95\columnwidth]{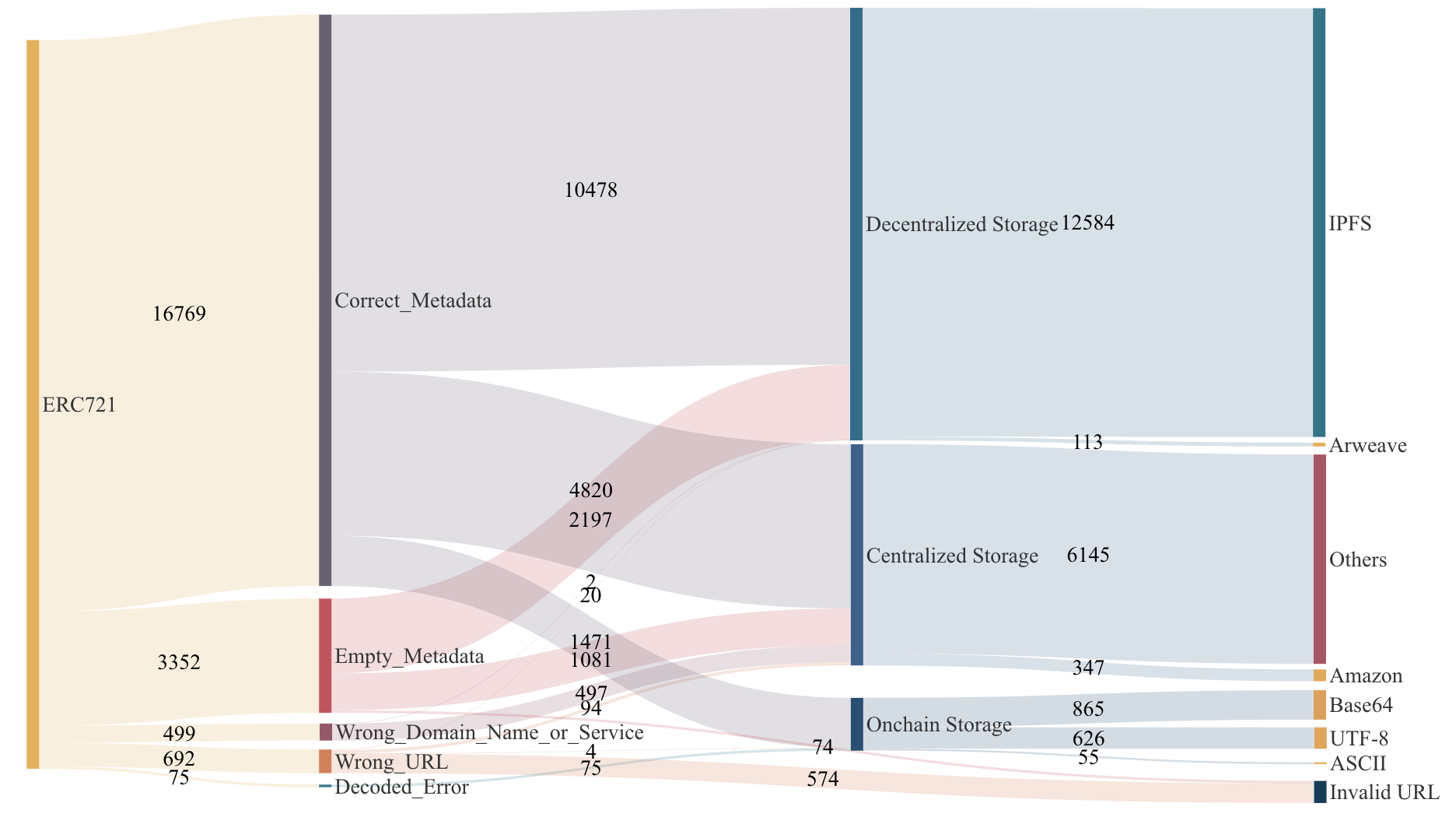}}
    \caption{NFT Metadata Details}
    \label{fig:NFT Metadata Details}
    \vspace{-5pt}
\end{figure}

\smallskip
\noindent\textbf{Contract addresses and tokens} (Figure \ref{fig:NFT Metadata Details}).
768,037 contract addresses are found in four token types: 334,533 ERC20 tokens with 43.5569\% of all contract addresses, 43,810 ERC721 tokens with 5.7042\% of, 3 ERC777 with 0.0004\% and 17,423 ERC1155 tokens with 2.2685\%.

\smallskip
\noindent\textbf{ERC721 analyses.}
We focus on the ERC721 data analysis due to its majority.
There are 43,810 ERC721 tokens which are NFTs. However, some metadata cannot be collected from the token URL successfully, due to empty metadata, wrong domain name or service, wrong URL and decoded error. There are only 21,387 metadata collected from the identiﬁed ERC721 tokens.

Acquired metadata refers to that metadata can be acquired from token URI and decoded from on-chain storage. There are 16,784 ERC721 token metadata that can be acquired correctly.

Empty metadata refers to that metadata can be acquired from token URI and decoded from on-chain storage, but the metadata is empty. The wrong domain name or service refers to the ERC721 token URI is not accessible due to the wrong domain name or service being unavailable. The wrong URL refers to the token URI is not correct and it is an invalid URI. The decoded error means that metadata is stored on-chain, but it cannot be decoded correctly. Figure \ref{fig:NFT Metadata Details} includes the count number for these four data types. 

There are three storage types: decentralized, centralized and on-chain.

\vspace{-5pt}
\begin{itemize}
\item 
For decentralized storage, IPFS and Arweave are two main storage solutions. 12,697 tokens are using a decentralized storage solution. 12,584 tokens are using IPFS to store their metadata and 113 tokens are using Arweave to store their metadata. 

\item
For centralized storage, Amazon is the well-known storage platform and 347 tokens store their metadata on the Amazon platform. The other 6145 tokens store their metadata on other diverse platforms, such as self-host storage solutions or other cloud providers' platforms.

\item
For on-chain storage, the metadata is encoded with different text encoding methods including Base64, UTF-8 and ASCII. 1,550 tokens encode their metadata and store the on-chain.
\end{itemize}

\begin{figure}[htbp]
\centering
    \resizebox{0.99\linewidth}{!}{
\begin{tikzpicture}
\begin{axis}[
    width=11cm,
    height=3cm,
    ybar,
    bar width=8pt,
    xlabel={Duplication Times},
    ylabel={Tokens},
    symbolic x coords={2,3,4,5,6,7,8,9,10,12,14,15,30,36,50},
    xtick=data,
    nodes near coords,
    nodes near coords align={vertical},
    ymin=0, ymax=750,
    xticklabel style={rotate=45, anchor=east},
    enlarge x limits=0.05
]
\addplot coordinates {(2,592) (3,81) (4,34) (5,18) (6,4) (7,5) (8,3) (9,3) (10,1) (12,1) (14,1) (15,1) (30,1) (36,1)};

\addplot[
    smooth, 
    color=red,
    line width=1.2pt,
    mark=none,
] coordinates {(2,592) (3,81) (4,34) (5,18) (6,4) (7,5) (8,3) (9,3) (10,1) (12,1) (14,1) (15,1) (30,1) (36,1)};
\end{axis}
\end{tikzpicture}
}
\caption{NFT Metadata Duplication}
\label{fig:nft-metadata-duplication}
    \vspace{-10pt}
\end{figure}
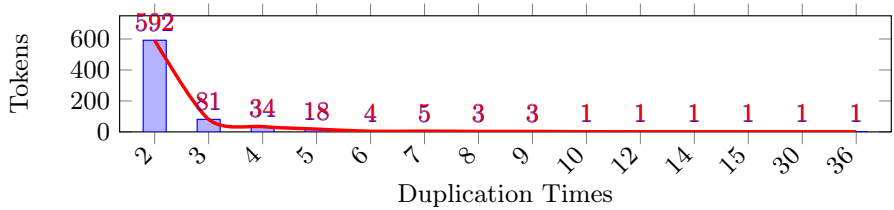

\smallskip
\noindent\textbf{Metadata duplication} (Figure \ref{fig:nft-metadata-duplication}).
In 16,784 of the ERC721 tokens metadata acquired successfully, there 14,466 metadata are not duplicated. 1,159 metadata are duplicated at different times.

\subsubsection{Artworks} 
\begin{figure}[htbp]
    \vspace{-20pt}
    \centering
    \centerline{\includegraphics[width=0.95\columnwidth]{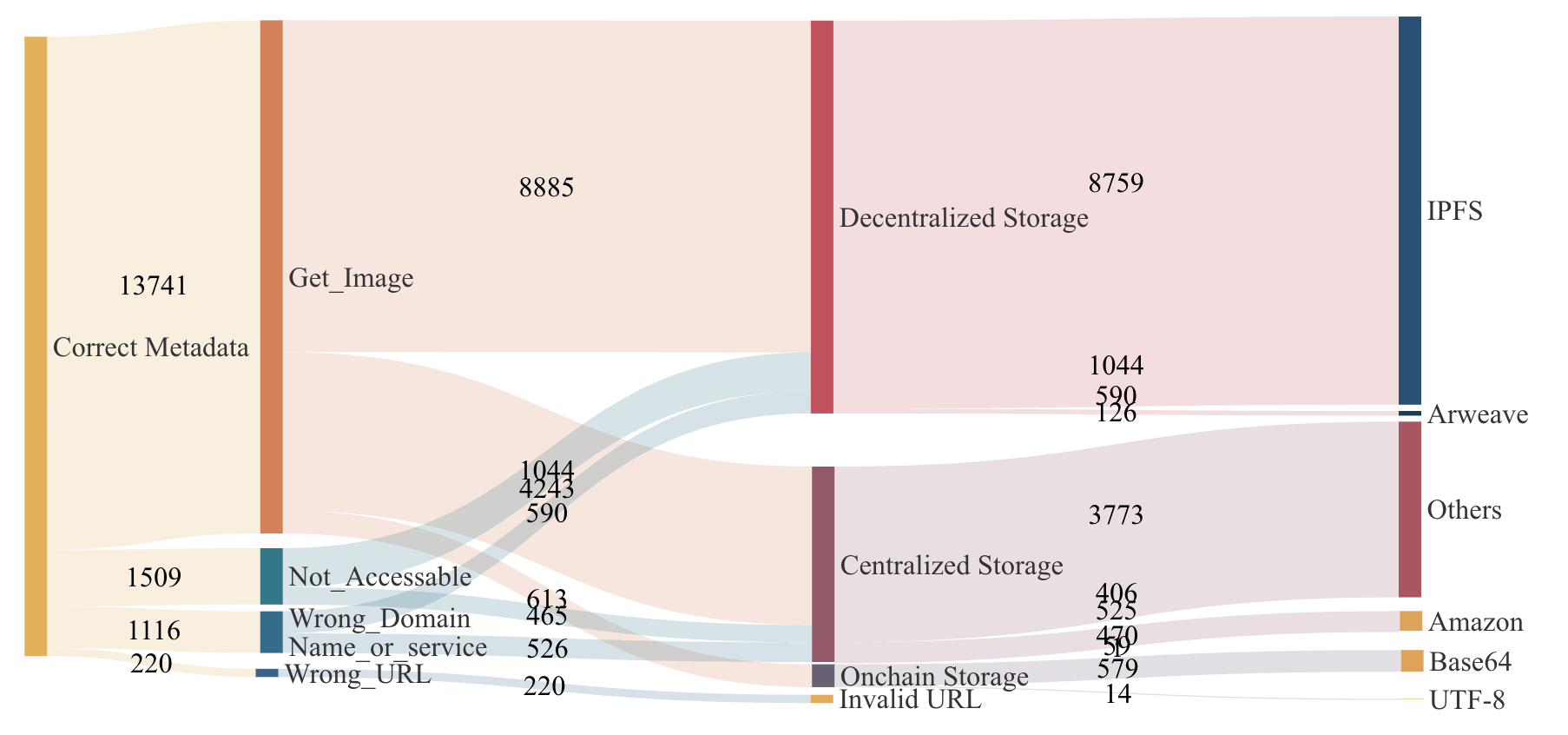}}
    \caption{NFT Artwork Details}
    \label{fig:NFT Artwork Details}
    \vspace{-20pt}
\end{figure}

(Figure~\ref{fig:NFT Artwork Details}).
The artwork data collection is from 16,784 acquired metadata. Among them, 57 pieces of metadata cannot be parsed as JSON objects correctly. 141 pieces do not have an image URL.

Other 16,586 metadata has image URLs. However, some images cannot be collected successfully due to being inaccessible, the wrong domain name or service, and the wrong URL. 13,741 artworks can be collected from metadata.

`Not accessible' means that the image URL can be accessed, but the acquired data from the image URL is not a correct image. A wrong domain name or service means that the image URL cannot be accessed due to the wrong domain name or service being down. The wrong URL means that the image URL is not in correct URL formats. Figure~\ref{fig:NFT Artwork Details} includes the count number for different data types. 

\vspace{-5pt}
\begin{itemize}
\item 
For artwork storage solutions, we have three types: decentralized storage, centralized storage, and on-chain storage. 

\item 
For decentralized storage, IPFS and Arweave are two main storage solutions. 10,519 artworks are using a decentralized storage solution. 10393 tokens are using IPFS to store their artworks and 126 artworks are stored on Arweave. 

\item 
For centralized storage, 530 tokens store their artwork on the Amazon platform. The other 4,704 tokens store their artworks on other platforms, such as self-host storage solutions or other cloud providers’ platforms. 

\end{itemize}
\begin{figure}[htbp]
    \vspace{-20pt}
    \centering
    \begin{minipage}[t]{0.4\textwidth}
        \centering
        \pgfplotstableread[col sep=comma,header=false]{
        Attribute,44.4,69
        In Description,42,26.8
        Under Attributes,3.7,4.2
        Under License,9.9,0
        }\data
        
        \pgfplotsset{
        percentage plot/.style={
            point meta=explicit,
            every node near coord/.append style={
            align=center,
            text width=1cm
        },
            nodes near coords={
            \pgfmathtruncatemacro\iszero{\originalvalue==0}
            \ifnum\iszero=0
                \pgfmathprintnumber{\originalvalue}$\%$\\
            \fi},
        nodes near coords align=vertical,
            yticklabel=\pgfmathprintnumber{\tick}\,$\%$,
            ymin=0,
            ymax=100,
            enlarge y limits={upper,value=0},
        visualization depends on={y \as \originalvalue}
        },
        percentage series/.style={
            table/y expr=\thisrow{#1},table/meta=#1
        }
        }
        \resizebox{0.9\linewidth}{!}{
        \begin{tikzpicture}
        \begin{axis}[
        axis on top,
        width=9cm,
        ylabel=Positions in Percent,
        percentage plot,
        ybar=0pt,
        bar width=0.75cm,
        enlarge x limits=0.25,
        symbolic x coords={Attribute, In Description, Under Attributes, Under License},
        xtick=data,
        xticklabel style={rotate=45, anchor=east}
        ]
        \addplot table [percentage series=1] {\data};
        \addplot table [percentage series=2] {\data};
        \legend{Copyright,License}
        \end{axis}
        \end{tikzpicture}
        }
        \caption{NFT License and Copyright Positions}
        \label{fig:NFT License and Copyright Positions}
    \end{minipage}
    \hfill
    \begin{minipage}[t]{0.58\textwidth}
        \centering
        \resizebox{\linewidth}{!}{
        \begin{tikzpicture}
        \pie[ color ={ blue!35, cyan!35, red!35, orange!35, green!35, pink!35, gray!35, teal!35 }]
        {18.99/CC0, 1.27/CBE-CC0, 45.57/CC BY-NC-ND 4.0, 1.27/CC-BY-4.0, 5.06/CC BY-NC-SA 4.0, 24.05/CC BY-NC 4.0, 2.53/CC BY NA 2.0, 1.27/CC BY-NC 2.0}
        \end{tikzpicture}
        }
        \caption{NFT License Standards}
        \label{fig: NFT License Standards}
    \end{minipage}
    \vspace{-10pt}
\end{figure}

For on-chain storage, the metadata is encoded with different text encoding methods including Base64, UTF-8 and ASCII. 613 tokens encode their artworks and store them on-chain.

\vspace{-10pt}
\subsubsection{Copyright.}
`Copyright' and `license' are two keywords to collect copyright information. By searching copyright as a keyword in all NFT metadata, there are 81 NFTs’ metadata that have copyright information. By searching keyword license, 168 NFTs’ metadata have license information. The copyright and license information are located in different places in metadata. Figure \ref{fig:NFT License and Copyright Positions} shows the 4 different places in metadata where copyright and license are located and how many NFTs are in every location respectively. 

\begin{table*}[htbp]
\vspace{-20pt}
\caption{NFT License and Copyright Types}
\begin{center}
\resizebox{\linewidth}{!}{
\begin{tabular}{c|c|l|l|c}
      \multicolumn{1}{c|}{\textbf{License Type}} & 
      \multicolumn{1}{c|}{\textbf{Description}} & 
      \multicolumn{1}{c|}{\textbf{Key Characteristics}} & 
      \multicolumn{1}{c|}{\textbf{Common Terms}} & 
      \multicolumn{1}{c}{\textbf{\, (\%)\, }} \\
      \midrule
    \makecell{Restricted Personal \\Use Licenses} & \makecell{Licenses that allow personal, \\non-commercial use with \\strict restrictions on rights. \\Copyright retained by creator.} & \makecell{- Non-commercial use only \\- No reproduction or modification \\- Copyright with creator \\- Marketplace display allowed} & \makecell{- Personal, non-commercial use \\- No reproduction \\- Copyright remains with creator} & 35\% \\ \hline
    \makecell{Commercial Use \\Licenses with \\Restrictions} & \makecell{Limited commercial use\\ with conditions. \\Copyright retained by creator.} & \makecell{- Limited commercial use \\(e.g., revenue caps) \\- Royalties on secondary sales \\- Copyright with creator \\- Specific usage permissions} & \makecell{- Limited commercial use \\- Revenue cap \\- Royalties} & 10\% \\ \hline
    \makecell{Creative Commons and \\Public Domain Licenses \\(CC0)} & \makecell{Broad use, \\including commercial, \\often under CC0. \\No copyright retention.} & \makecell{- Full use, e.g., commercial/derivatives \\- Copyright waived or transferred \\- No restrictions} & \makecell{- CC0 \\- Public domain \\- No rights reserved} & 15\% \\ \hline
    \makecell{Custom or \\Proprietary Licenses} & \makecell{Bespoke licenses tailored \\to specific NFT projects.} & \makecell{- Project-specific terms \\ \makecell{- Unique benefits \\(e.g., event access)} \\- Copyright with creator} & \makecell{- Refer to external guidelines \\ \makecell{- Unique benefits \\(e.g., event access)} \\- Creator retains copyright} & 28\% \\ \hline
    \makecell{Creative Commons \\Licenses (Non-CC0)} & \makecell{Licenses using Creative \\Commons frameworks \\(e.g., CC BY-NC-SA) \\with conditions like \\attribution or \\non-commercial use.} & \makecell{- Attribution required \\- Non-commercial  conditions \\- Copyright with creator} & \makecell{- Attribution required \\- Non-commercial \\- Share-alike} & 20\% \\ \hline
    No License Specified & \makecell{No explicit license terms, \\creating ambiguity.} & \makecell{- No clear usage rights \\- Copyright with creator by default \\- Ambiguous holder permissions} & \makecell{- Copyright © [Year] [Creator] \\- No license terms provided} & 5\% \\ \hline
    Specialized Licenses & \makecell{Unique licenses addressing \\specific NFT types \\(e.g., fonts) or ethical \\considerations \\(e.g., hate speech).} & \makecell{- Tailored to NFT type \\(e.g., NFTypeface) \\- May include ethical terms \\- Copyright with creator} & \makecell{- Can’t Be Evil \\- NFTypeface \\- Ethical restrictions} & 2\% \\
\end{tabular}
}
\label{tab: NFT License and Copyright Types}
\end{center}
\vspace{-20pt}
\end{table*}

Next, the 81 NFTs’ metadata that have the copyright information and 168 NFTs’ metadata that have license information are merged together as one file with 203 NFTs’ metadata. There are 79 NFTs’ metadata that have standard license information. Figure \ref{fig: NFT License Standards} shows the distribution for every standard license. The other 124 NFTs’ metadata does not have standard license information. However, some of them have custom copyright information. Table \ref{tab: NFT License and Copyright Types} shows all 203 NFTs’ license and copyright types, including license type, description, key characteristics, common terms and percentage.

\smallskip
\noindent\textbf{Artists identification.}
There is an artist name list with 16812 artists from the Australian Copyright Agency \footnote{Australian Copyright Agency: \url{https://www.copyright.com.au}}. This artist's name list is used to carry out identification from NFT metadata. Based on the correct metadata, the identification is done. There are 199 articles identified from the 13462 correct metadata, which is 1.4782\% of all correct NFT metadata (cf. Table~\ref{tab: Artists Identification}). The name is listed in the table. However, some artists' names are incorrect.
There are 8 NFTs found by search keyword: Australia in metadata, but only one NFT has the artist's name (cf. Table~\ref{tab: Australia in Metadata}). And 2 NFTs’ artwork names are duplicated. The token address, artwork name, and artist are shown in the table. 
\begin{table*}[htbp]
    \vspace{-15pt}
    \centering
    \begin{minipage}[t]{0.4\textwidth}
        \centering
        \caption{Artists Identification}
        \resizebox{1\linewidth}{!}{
        \begin{tabular}{c|c|c}
            \textbf{Name} & \textbf{Total numbers} & \textbf{Percentage} \\
            \midrule
             & 199 & \\
            What & 103 & 51.7588\% \\
            Bandi & 18 & 17.4757\% \\
            Rose & 45 & 22.6131\% \\
            Green Jack & 6 & 3.0151\% \\
            ME & 10 & 5.0251\% \\
            Yama & 4 & 2.0101\% \\
            Pcd & 4 & 2.0101\% \\
            Mama & 3 & 1.5075\% \\
            Ipg & 3 & 1.5075\% \\
            Nell & 2 & 1.0050\% \\
        \end{tabular}
        }
        \label{tab: Artists Identification}
    \end{minipage}
    \hfill
    \begin{minipage}[t]{0.58\textwidth}
        \centering
        \caption{Australia in Metadata}
        \resizebox{1\linewidth}{!}{
        \begin{tabular}{c|c|c}
            \textbf{Address} & \textbf{Artwork Name} & \textbf{Artist Name} \\
            \midrule
            0x1eD1...c7c5 & Aidan Lee - Australia & \\
            0xbB39...7F9B & Do Not Shoot Me \#2 & \\
            0x0ea8...3777 & The dreamer & Lilyillo \\
            0x22C1...88d5 & Seby Round The World & \\
            0xfa77...EF26 & Do Not Shoot Me \#2 & \\
            0xF6ef...3399 & Australian white breeding sheep 01 & \\
            0xdBd...3F80 & Australian White Breeding Sheep 01 & \\
            0x1dC3...8F35 & Blue Hour Adventures \#0 & \\
        \end{tabular}
        }
        \label{tab: Australia in Metadata}
    \end{minipage}
    \vspace{-30pt}
\end{table*}

\section{Analysis and Discussion}
\subsection{Copyright Issues}
The total number of acquired NFT metadata is 16769, but only 203 NFTs have copyright and license information, of which 1.21\% of NFT concern the copyright protection. There are 4 locations for copyright and license in metadata, which means there is no standard location for copyright or license in metadata. The high frequency of `license' as an attribute (69.0\%) indicates that creators prioritize structured, machine-readable fields for licensing terms, enhancing clarity for platforms and users. Copyright appears less frequently (81 occurrences) and is split almost evenly between attributes (44.4\%) and descriptions (42.0\%). This suggests that copyright declarations are less standardized and often supplementary. In addition,  both `copyright' (3.7\%) and `license' (4.2\%) have minimal representation under OpenSea’s standardized attributes, despite the dominance of structured attributes for `license' (69.0\%). This suggests that while creators value structured metadata, they often use custom or non-standard fields, which may reduce interoperability across platforms. Both `copyright' (3.7\%) and `license' (4.2\%) have minimal representation under OpenSea’s standardized attributes, despite the dominance of structured attributes for `license' (69.0\%). This suggests that while creators value structured metadata, they often use custom or non-standard fields, which may reduce interoperability across platforms. Restricted Personal Use Licenses (35\%) and Custom or Proprietary Licenses (28\%) together account for 63\% of the dataset, highlighting a strong preference for licenses that maintain creator control over intellectual property. This means that creators seek to protect their work while offering limited usage rights to holders. Creative Commons-Based Licenses (Non-CC0) (20\%) and Creative Commons and Public Domain (CC0) (15\%) together represent 35\% of the dataset, indicating widespread use of Creative Commons frameworks. Commercial Use Licenses with Restrictions (10\%) are relatively uncommon, which permit limited commercial use with royalties. 

The low standardization is particularly notable for copyright, which is less structured than licensing terms, potentially complicating ownership verification for users. For creators, this means they are good at using attributes for licenses but need to standardize more, especially for copyright, which often hides in descriptions. Moreover, Creators should adopt explicit licenses, even for restrictive terms. Users benefit from clear license attributes but might struggle with copyright details in free-text fields, and platforms could help by pushing for standards or better tools to show this info. Creators should move legal stuff to structured attributes, keep copyright and license separate, and follow OpenSea’s format to make things smoother. Platforms should also offer guides and features like license badges to make metadata easier to understand. Additionally, platforms could develop tools or templates for creators to define clear commercial terms (e.g., revenue caps, royalties), encouraging monetization while protecting intellectual property. For Custom Licenses (28\%), platforms should improve the display of external guidelines to ensure holders can easily access and understand project-specific conditions. By adopting structured metadata, separating copyright and license, and leveraging platform tools, the NFT ecosystem can enhance transparency and usability.

\begin{center}
\fbox{%
    \begin{minipage}{0.9\linewidth}

    \textbf{Answer to RQ1.} \par
      The copyright issues include less concern about copyright protection and Less copyright and license standardization in NFT metadata.
 
    \end{minipage}
}
\vspace{-10pt}
\end{center}

\subsection{Security Issues}
\begin{figure*}[htp]
    \vspace{-10pt}
    \centering
    \includegraphics[width=\linewidth]{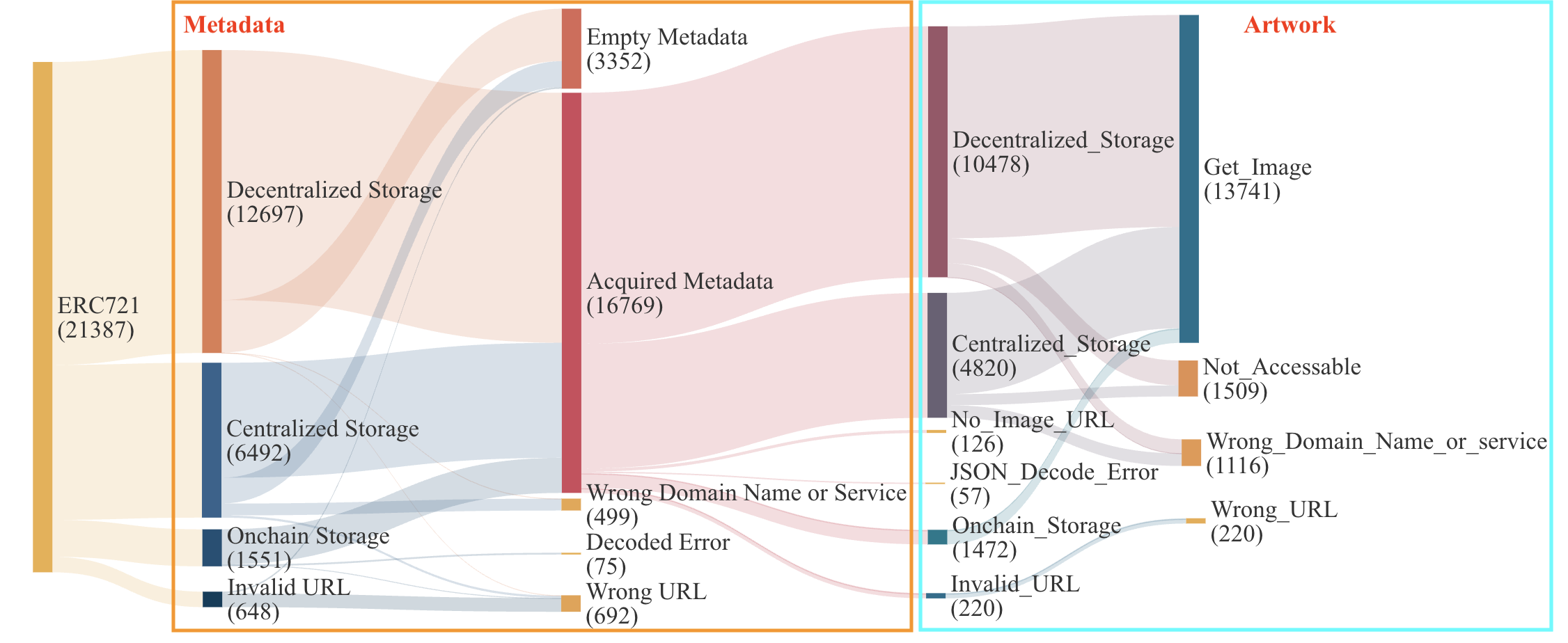}
    \caption{NFT Metadata and Artwork Details}
    \label{fig:NFT Metadata and Artwork Details}
    \vspace{-20pt}
\end{figure*}

The provided data highlights several security challenges associated with NFT metadata and artwork storage, including risks of metadata tampering, artwork tampering, missing metadata pointers, and missing artwork pointers. Decentralized storage shows a high volume of acquired metadata (10,478), but the presence of empty metadata (2,197) and inaccessible images (914) poses significant risks. Missing or incorrect metadata pointers, such as wrong domain names or services (570 cases in decentralized storage), can lead to tampering vulnerabilities, where metadata may be altered or replaced, undermining the integrity of the NFT. Similarly, centralized storage, though effective in acquiring metadata (4,820), is highly susceptible to security risks due to its reliance on third-party servers. The presence of wrong URLs (94) and inaccessible images (337) indicates that metadata and artwork pointers are prone to failure, potentially allowing malicious actors to alter or delete data. On-chain storage, while the most secure in terms of immutability, has limited adoption (1,472 total metadata acquired) due to high costs. However, it demonstrates strong resistance to tampering, with no reported issues of wrong domain names or inaccessible images in its artwork storage. Invalid URLs, on the other hand, represent the most significant vulnerability, with 574 instances of wrong URLs and 74 cases of empty metadata, emphasizing poorly implemented NFTs that lack robust pointers for both metadata and artwork. These missing pointers can lead to loss of ownership proof and expose NFTs to exploitation. Overall, the data underscores the importance of adopting secure storage practices, such as on-chain storage for critical assets and regular audits to detect and address missing or incorrect metadata and artwork pointers, ensuring the integrity of NFTs. Figure \ref{fig:NFT Metadata and Artwork Details} provides more details.

In terms of security, addressing metadata and artwork pointer vulnerabilities should be a priority for future research. Decentralized storage systems face notable challenges, such as tampering the risks caused by missing or incorrect pointers and centralized storage depends heavily on third-party servers, increasing risks of data breaches. On-chain storage offers strong tamper resistance but is limited by high costs. Future studies should explore cost-effective strategies to enhance security, such as partial on-chain data storage for key pointers or metadata hashing mechanisms to detect unauthorized modifications. Moreover, developing better validation and auditing tools to address errors in URLs and metadata can improve the overall reliability of NFTs.

\begin{center}
\fbox{%
    \begin{minipage}{0.9\linewidth}

    \textbf{Answer to RQ2.} \par
      The metadata tampering, artwork tampering, missing metadata pointers, and missing artwork pointers are NFT security issues.
 
    \end{minipage}
}
\vspace{-10pt}
\end{center}

\subsection{Artists Identification}
According to the artists' identification result, it is hard to identify the Australian artists from the NFT metadata. There are some reasons perceived from the metadata collection and analysis.

\textbf{1. No geographical information in NFT metadata.}

Most NFT metadata uses  Opensea metadata standard, but this metadata standard does not include geographical information. It causes difficulty in identifying the NFT geographical location or country.

\textbf{2. The different artists' names in the NFT metadata.}

The names in the artist's name list are artists' full names. However, in the NFT metadata, they use a different name as the author, such as art name, short name or nickname. For example, Attafuah, Serwah Ama Gyekyewah Bianca is the name of one single artist in the list. However, in her NFT metadata, the real name is Serwah Attafuah which is different from her full name.

\textbf{3. The issues from the artist's name list.}

Two issues arise from the artist's name list. Firstly, the list does not include all Australian artists. Some artists do not register in the Australian Copyright Agency system. Thus, their name is not in the artist's name list. Second, the list has some incorrect names, such as `What', `ME', etc.

\smallskip
Identifying artists from NFT metadata presents unique challenges due to the lack of geographical data and inconsistencies in name representation. Future research should explore enriching metadata standards to include optional fields like artist nationality or residence, which would aid in categorizing NFTs geographically. Additionally, improving artist databases to address issues like incomplete or incorrect entries and incorporating mechanisms to link pseudonyms or nicknames with real names could enhance identification accuracy. Collaborative efforts between artists, NFT platforms, and copyright agencies will be essential to streamline artist recognition in the NFT ecosystem. Moreover, utilizing machine learning algorithms to cross-reference artist metadata with existing databases might provide a scalable solution for tracing artists to their NFTs.

\begin{center}
\fbox{%
    \begin{minipage}{0.9\linewidth}

    \textbf{Answer to RQ3.} \par
      It is very hard to identify artists from NFT metadata, due to different reasons: no geographical data, inconsistencies artists' name, inconsistencies in name artists name list and wrong name in artists name list.
 
    \end{minipage}
}
\end{center}

Future research should address the interconnected challenges of copyright, security, and artist identification to build a more accessible NFT ecosystem. By combining technical advancements with policy reforms and education, the NFT space can evolve into an artist-friendly environment.

\section{Conclusion}
\label{sec-conclusion}
\vspace{-5pt}
In this paper, we collected NFT metadata and artwork images from on-chain storage, centralized storage, and decentralized storage. The result of the collected data is categorized into four types: metadata, artwork, copyright, and artists. Our analysis and discussion focus on three key aspects: copyright issues, security issues, and artist identification. The analysis reveals several challenges. In terms of copyright, there is limited attention to copyright protection and a lack of standardized copyright and license information in NFT metadata. Security issues include metadata tampering, artwork tampering, missing metadata pointers, and missing artwork pointers. For artist identification, we observed missing geographical data, inconsistencies in artist names, and mismatches between name lists and actual artist identifiers. These challenges indicate that NFTs face significant limitations in effectively safeguarding the copyright and security of artworks, particularly for Australian and Indigenous artists. Moreover, it is not yet practical for the Australian Copyright Agency to accurate artist identification from NFT metadata. Addressing these issues in the future will enhance the integrity and reliability of NFTs, making them a more valuable asset that bridges Australian digital artwork and the decentralized ecosystem.

\vspace{-10pt}
{\footnotesize \bibliographystyle{splncs04}
\bibliography{bib}}

\end{document}